# Relationship between Gender and Code Reading Speed in Software Development

Yuriko Takatsuka
Kindai University
HigashiOsaka, Japan

Yukasa Murakami
Kindai University
HigashiOsaka, Japan

Masateru Tsunoda
Kindai University
HigashiOsaka, Japan
tsunoda@info.kindai.ac.jp

Masahide Nakamura
Kobe University
Kobe, Japan
masa-n@cs.kobe-u.ac.jp

## ABSTRACT

Recently, workforce shortage has become a popular issue in information technology (IT). One solution to increasing the workforce supply is to increase the number of female IT professionals. This is because there is gender imbalance in information technology area. To accomplish this, it is important to suppress the influence of biases, such as the belief that men are more suited for careers in science and technology than women, and to increase the choice of careers available to female professionals. To help suppress the influence of gender bias, we analyzed the relationship between gender and code reading speed in the field of software development. Certain source codes require developers to use substantial memory to properly understand them, such as those with many variables that frequently change values. Several studies have indicated that the performance of memory differs in males and females. To test the veracity of this claim, we analyzed the influence of gender on code-reading speed through an experiment. Pursuant to this, we prepared four programs that required varied amounts of memory to properly understand them. Then, we measured the time required by each of the 17 male and 16 female subjects (33 subjects in total) to comprehend the different programs. The results suggest that there is no explicit difference between male and female subjects in this regard, even in the case of programs that require high memory capacities for proper understanding.

## CCS CONCEPTS

Software and its engineering → Software creation and management → Collaboration in software development → Programming teams

## KEYWORDS

Developer performance, code reading, mental simulation

## 1 Introduction

Recently, software has become a widely used infrastructure in society, and consequently, workforce shortage in information technology (IT) has become a popular issue. For example, the White House blog [20] indicated a huge gap between workforce demand and supply in the Unites States (US) based on the data published by the Bureau of Labor Statistics. Similarly, there is a similar gap in Europe [15]. Further, TechRepublic [28] stated that the shortage of technological talent continues to be a problem for many CIOs.

One solution to the issue is to increase the workforce supply by increasing the number of female IT professionals [1], [15], which is low because of the gender imbalance in the field. For instance, in 2018, although 57% of all professionals in the US were reported to be female, merely 26% of IT professionals were female [26]. Europe also suffers from similar gender imbalance in the field of IT, as indicated by the data published by Eurostat [15]. Therefore, there is room for improvement in the percentage of female IT professionals, considering the percentage of total female professionals. Additionally, increasing the number of the former is beneficial to both male and female employees. For example, Hoogendoorn [18] found that business teams with balanced gender composition performed better (e.g., better sales and profit) than teams with biased gender composition.

To increase the number of female IT professionals, it is important to suppress the influence of patriarchal biases, such as the belief that men are more suited for careers in science and technology than women, and to increase the choice of careers available to female professionals [6]. Further, it is important to eliminate barriers to women's achievements. For example, 46% of female professionals who work in Fortune 1000 companies indicated that gender-based stereotypes is a barrier to their professional advancement [8]. One method to address the issue is to present empirical evidence that contradicts such biases. In the field of medicine, several studies have been conducted that provide such evidence. For example, Tsugawa et al. [34] compared patient outcomes, or prognoses, when treated by female physicians to that by male physicians, and demonstrated that the performance of female physicians was better.

To suppress the influence of gender bias on IT professions, we analyzed the relationship between gender and code reading speed in software development. Code reading is one of important activities related to software development. When we modify source codes or review them, we have to read them to understand them. About 90% of software life cost is related to its maintenance phase [10], and modifying source codes (and reading them) is needed on software maintenance. Speed of software development is one of important factors [5]. Some developers think that the most important developer performance metrics is speed of developer [35].

Certain source codes may require developers to use substantial memory to achieve complete understanding of them [25], such as those with several variables that frequently change values. Many studies have indicated that memory performance differs in males and females [21]. Thus, this study was started with the assumption that memory may be impacted by gender. To quantitatively measure the extent of memory required by an individual to understand a program completely, we employed understandability metrics proposed in prior studies [19][25]. These metrics evaluate understandability based on the amount of memory needed to understand a program, instead of the complexity of the program. In other words, when a program is considered to be difficult to understand by the metrics, it indicates that substantial memory is required to achieve proper understanding. In the experiment, we compared the speeds with which male and female developers understood such programs.

Several studies have been conducted with the aim to support female IT developers. For example, May et al. [23] analyzed the data on rewards and answers on the website Stack Overflow focusing on the gender of the developers, and concluded that the behavior of developers varied by gender. In the study, the authors extracted data from Stack Overflow and aimed to improve the system design to encourage participation from female IT developers. Gorbacheva et al. [15] proposed six directions for future research on gender imbalance among IT professionals, based on literature reviews and their own practical experiences. However, to our knowledge, no existing research has directly examined the relationship between the gender of the developer and code reading speed based on subjective experiments, to address the gender bias in IT.

Some studies have suggested that correcting the gender bias has direct advantages. For example, Spencer et al. [31] termed the harmful influence of gender bias a "stereotype threat." Using subjective experiments, they showed that the difference in performances in mathematics tests could be eliminated by suppressing the stereotype. This finding has been corroborated by other studies [3]. Therefore, our study is also expected to have such an advantage when our results are utilized to correct the gender bias in IT.

## 2 Related Work

Memory is required to properly understand source code because the code relies on many variables to implement its various functions. Therefore, memory performance directly influences code reading performance. In the field of psychology, many studies demonstrate that memory performance differs between males and females [21]. For example, Herlitz et al. [16] confirmed that males exhibit stronger spatial memory, while females demonstrated better verbal memory. They also suggested that the advantage of the enhanced verbal memory performance of females is applicable in many different situations. Fuentes et al. [13] showed that females recall more episodic information than males. Pauls et al. [27] determined that females outperform males in auditory memory. Total immediate recall of word lists by females was greater than that of males in experiments performed by Sunderaraman et al. [32]. Maitland et al. [22] found that the declarative memory, which consists of episodic and semantic memory, of females was also superior to that of their male peers. Based on these studies, this study also focused on the impact of developer gender on code reading.

To support women software developers, other studies have analyzed the differences in behavior based on gender. For example, Terrell et al. [33] compared acceptance rate of pull requests on open source software for female developers versus male developers. They found that females' contributions tended to be accepted more often than those of males. The result indirectly suggests that female developers may be more competent on OSS projects, and it is effective to suppress the influence of the gender biases. To improve software effectiveness for females, some studies analyzed differences of software feature usage by gender. For instance, Burnett et al. [7] used questionnaires to analyze differences of software feature usage by gender. They concluded that there were meaningful differences between genders. For the similar goal, Beckwith et al. [2] analyzed gender differences for end users in the context of debugging spreadsheets by subjective experiments. To support career choice of females, Beyer et al. [4] analyzed gender differences among students who majored in Computer Science (CS). They found that male students had greater confidence in using computers than female students. However, code reading speed was not directly evaluated by subjective experiments in the study, and they did not focus on developers' memory performance.

In other fields, additional studies have also focused on performance differences attributed to gender. For example, in the field of psychology, Chabris et al. [9] examined player performance in the game of chess, and sought to determine why male players are dominant in elite chess. Using a cohort study, they concluded that greater numbers of boys entering chess at the lowest levels caused the dominance of male players.

## 3 Program Understandability Metrics

To measure program understandability, past studies [19][25] have proposed metrics based on human memory. The metrics assume that mental simulation [11] is applied to understand a program. Mental simulation refers to when a developer comprehends a program without external aids such as computers or memos, that is, they rely only on their own thoughts. Mental simulation is

```
int i , t;            int i , t;

t = 11;               t = 11;
t = t - 1;            t = t - 1;
i = 2;                i = 2;
if(i < t){            if(i < t){
    i = i + 2;            t = t - 2;
    if(i < t){            if(i < t){
        i = i + 2;            i = i + 2;
    }                     }
    if(i < t){            if(i < t){
        i = i + 2;            t = t - 2;
        if(i < t){            if(i < t){
            i = i + 2;            i = i + 2;
        }                     }
    }                     }
    if(i < t){            if(i < t){
        i = i + 2;            i = i + 2;
    }                     }
}                     }
System.out.println("i = " System.out.println("i = "
+i);                  +i);

         (a0)                  (a1)
```

**Figure 1: Program a0 and a1 [19]**

```
int a, b, c, d, e, f, g;   int a, b, c, d, e, f, g;

a = 2;                     a = 2;
b = 4;                     b = 4;
c = 3;                     c = 3;
d = 6;                     d = 6;

c = c + 4;                 c = c + 4;
d = d - 2;                 d = d - 2;
if(c < 5)                  a = a * 2;
    e = d + 5;             b = b + 6;
else                       if(a > 7)
    e = d + 3;                 f = b - 3;
a = a * 2;                 else
b = b + 6;                     f = b - 5;
if(a > 7)                  if(c < 5)
    f = b - 3;                 e = d + 5;
else                       else
    f = b - 5;                 e = d + 3;
g = e + f;                 g = e + f;

System.out.println("g = "+g); System.out.println("g = "+g);

         (b0)                  (b1)
```

**Figure 2: Program b0 and b1 [19]**

often applied when a developer reads relatively small code fragments. When applying mental simulation, values of variables must be remembered by the developer. However, it is difficult to accurately recall numerous values. So, the studies [19][25] assume that cost of mental simulation is high for a program where many variable values must be remembered in order to properly understand the code. Thus, the understandability of that same program is regarded as low.

In the study [25], short-term memory is regarded as a first in, first out (FIFO) queue, and they comprise the virtual mental simulation model (VMSM). Understandability metrics were then created based on the VMSM. The basic idea of the metrics is that the size of FIFO queue is limited. When a developer refers to a variable and its value that are stored in the queue, the cost of the mental simulation is regarded as low. In contrast, when the value is not stored in the queue, the cost is regarded as high. This is because the developer has to backtrack to the point where the value of the variable was changed. In the study, four metrics were defined as follows:

- ASSIGN: cost with regard to variable assignment;
- RCL: cost of recalling a value of a variable in short-term memory;

**Table 1. Understandability of programs used in the experiment [19]**

| Program | ASSIGN | RCL | BT_CONST | BT_VAR | SUM_UPD | VAR_UPD |
|---------|--------|-----|----------|--------|---------|---------|
| a0      | 12     | 6   | 0        | 80     | 7       | 1.25    |
| a1      | 12     | 6   | 0        | 48     | 7       | 0.25    |
| b0      | 18     | 8   | 1        | 30     | 11      | 0.24    |
| b1      | 18     | 6   | 1        | 54     | 11      | 0.24    |

- BT_CONST: cost of acquiring the value of a constant as the number of times developer backtracks to find the constant;
- BT_VAR: cost of acquiring the value of a variable in terms of the distance of backtracking to the variable.

The study [19] noted that the metrics do not consider recalculation cost based on the number of times about updating variables. Furthermore, the study observed that the backtracking metrics are compromised in that they are over-sensitive to changes in the order of the lines of code. To solve the problem, the study proposed two metrics about understandability. First, the number of updates of each variable are regarded as elements of a vector. The metric, named SUM_UPD, is based on the sum of the elements. Second, the authors defined another metric, SUM_VAR, based on the variance of the elements.

## 4 Experiment

### 4.1 Overview

The purpose of this experiment is to clarify whether developers' gender affects code reading speed or not, especially when developers read a program which needs more memory to understand. To analyze that, we prepared programs which require different amounts of memory for proper understanding. Thereafter, we measured the time subjects required to understand the programs.

We used four programs shown in the study [19] as the programs with varying memory requirements. We referred to them as program a0, a1, b0, and b1. They are depicted in Figures 1 and 2. The size of each program was approximately 20 to 30 lines of code. We asked subjects that what the value of a variable would be after the program executed. If their answer was correct, the subject was regarded as understanding the program correctly. For example, for program a0, subjects were asked to provided the value of the variable "i" after execution. The subjects had to achieve understanding of the programs based on the mental simulation alone. They were not allowed to use memos or other external aids.

To measure the amount of memory needed to understand the programs, we used the six metrics explained in Section 3 (i.e., ASSIGN, RCL, BT_CONST, BT_VAR, SUM_UPD, VAR_UPD). Table 1 details the understandability of the programs based on the metrics. Note that the result refers to the study [19]. Based on the metrics ASSIGN, BT_CONST, and SUM_UPD, programs b0 and b1 need substantially more memory for proper understanding compared to programs a0 and a1. Additionally, RCL indicates that program b0 needs most memory to understand it.

We classified subjects into groups according to gender, that is, a male group and a female group, and calculated statistics such as average and median time to answer. After that, we compared the difference between the male group and the female group. Subjects were undergraduate students who study computer science at the same university. The total number of subjects was 33, of which 17 were male and 16 were female. Their ages were similar and they all shared the same nationality.

To avoid the influence of the order of reading programs, we changed the order for each subject in the experiment. For example, one subject read the programs in the order of program a0, a1, b0, and b1. Another subject read the programs in the order of program b1, a0, b0, and a1. In the experiment, we assumed that a program which needs more memory is more quickly understood by female developers than their male counterparts. Note that the assumption is a sort of null hypothesis. Therefore, we can interchange male and female in the assumption and the following research questions. We set three research questions as follows:

- RQ1: Comparing the male and female groups, which group was faster with regard to understanding programs?
- RQ2: When subject read memory intensive programs, did the female group realize faster understanding speeds than did the male group?
- RQ3: When subjects of female group read programs, was the difference of understanding speed small between programs requiring much memory and programs requiring less memory?

RQ1 was set to analyze the difference of code reading speed between genders simply. RQ2 was set to analyze code reading speed on under the specific condition (i.e., reading read memory intensive programs). RQ3 was set to analyze the influence of read memory intensive programs, focusing on intragroup (i.e., on each gender), whereas RQ2 focuses on intergroup.

**Table 2. Basic statistics of subjects**

| | | Average | Median | SD |
|---|---|---|---|---|
| Male | Age | 21.3 | 21 | 0.7 |
| | Years of experience | 3.6 | 3 | 1.9 |
| Female | Age | 21.5 | 21 | 1.2 |
| | Years of experience | 2.8 | 3 | 0.8 |

**Table 3. Correlation coefficients to reading time**

| | Correlation coefficient | p-value |
|---|---|---|
| Age | 0.08 | 0.38 |
| Years of experience | 0.04 | 0.61 |
| ASSIGN | 0.36 | **0.00** |
| RCL | 0.13 | 0.13 |
| BT_CONST | 0.36 | **0.00** |
| BT_VAR | -0.29 | **0.00** |
| SUM_UPD | 0.36 | **0.00** |
| VAR_UPD | -0.44 | **0.00** |

**Table 4. Answering time stratified sex (seconds)**

| | | a0 | a1 | b0 | b1 |
|---|---|---|---|---|---|
| Male | Average | 51.9 | 87.6 | 85.0 | 132.9 |
| | Median | 54 | 78 | 75 | 89 |
| | SD | 20.8 | 48.9 | 42.7 | 106.0 |
| Female | Average | 56.5 | 95.6 | 130.1 | 106.2 |
| | Median | 52.5 | 83 | 120.5 | 110 |
| | SD | 17.1 | 57.4 | 89.7 | 38.3 |
| | p-value | 0.29 | 0.63 | 0.12 | 0.87 |

**Table 5. Ratio of answering time to program a0**

| | | a1 / a0 | b0 / a0 | b1 / a0 |
|---|---|---|---|---|
| Male | Average | 1.84 | 1.79 | 2.93 |
| | Median | 1.31 | 1.59 | 2.03 |
| | SD | 1.24 | 1.01 | 2.88 |
| Female | Average | 1.69 | 2.49 | 2.09 |
| | Median | 1.37 | 1.75 | 1.97 |
| | SD | 0.78 | 1.83 | 1.06 |
| | p-value | 0.87 | 0.44 | 0.58 |

## 4.2 Tool for the Experiment

We developed a tool for the experiment. It shows codes of programs for subjects, and measures answering time and the number of incorrect answers. The tool was made using macros on spreadsheet software. The behavior of the tool is as follows:

1. Both the program code and text box dialog are shown when a subject selects the "Answer" button.

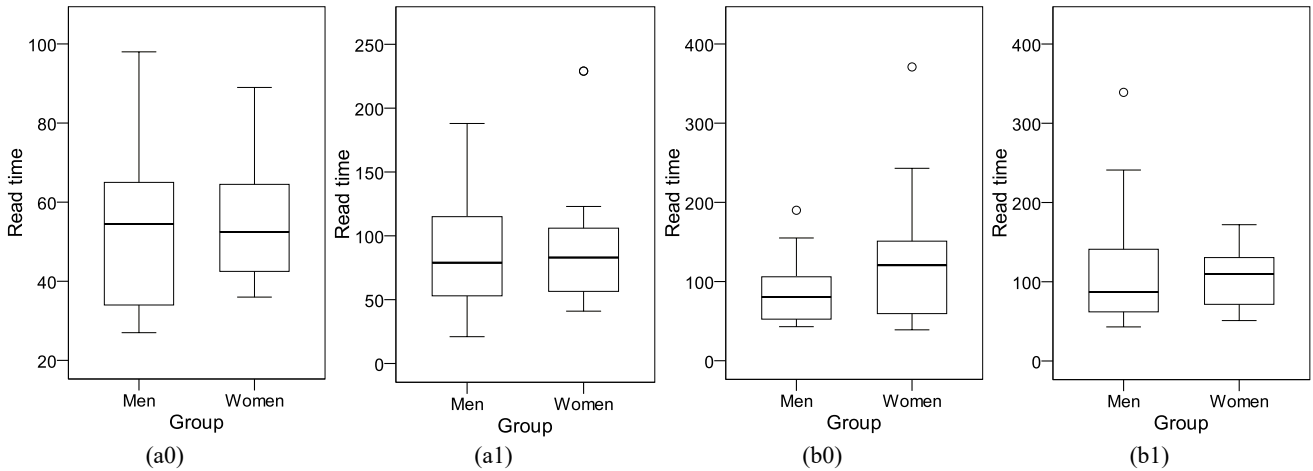

**Figure 3: Answering time of each program**

2. The text box is displayed until the subject replies the correct answer.
3. After the code of the program is shown, the tool records answering time and the number of incorrect answers. These values are also shown to the subject during the experiment.
4. When the subject replies the correct answer, the tool shows next program.

### 4.3 Result

**Influence of programming experience**: Table 2 shows basic statistics of subjects. In the tables, SD means standard deviation. Although years of experience of male group was slightly longer than females, the medians were similar. Age was also similar. Table 3 shows relationships between answering time and metrics. We used Spearman's rank correlation coefficient to avoid the influence of outliers. As shown in the table, age and years of experience did not affect code answering time. On ASSIGN, BT_CONST, and SUM_UPD were positively correlated with the answering time, and the correlation coefficients were significant at a level of 0.05. Therefore, these metrics effectively demonstrate the difficulty of code reading.

**Answer to RQ1 and 2**: Table 4 shows the average, median, and standard deviation of answering times for the male groups and female groups. When focusing on the average, the times of the male group were shorter than those of the female group, except for program b1. When focusing on the median, the time of the male group was shorter than the female group, except for program a1 and a0. However, the differences were not large, nor were the differences statistically significant on 0.05 level (see the bottom row of Table 4), when we used Mann-Whitney U test.

Distributions of the answering times for each group are shown in Figure 3 using boxplots. Overall, the distributions of answering times showed little difference among programs, except for program b0. Interquartile range of the times of the female group was slightly smaller than the male group, again excluding program b0. Therefore, we did not conclude that either group achieved shorter answering times. So, the answer to RQ1 is, "Program understanding speed of male developers was similar to that of female developers." The answer to RQ2 is, "No. There are not explicit differences in the understanding times of male and female developers when they read memory-intensive programs."

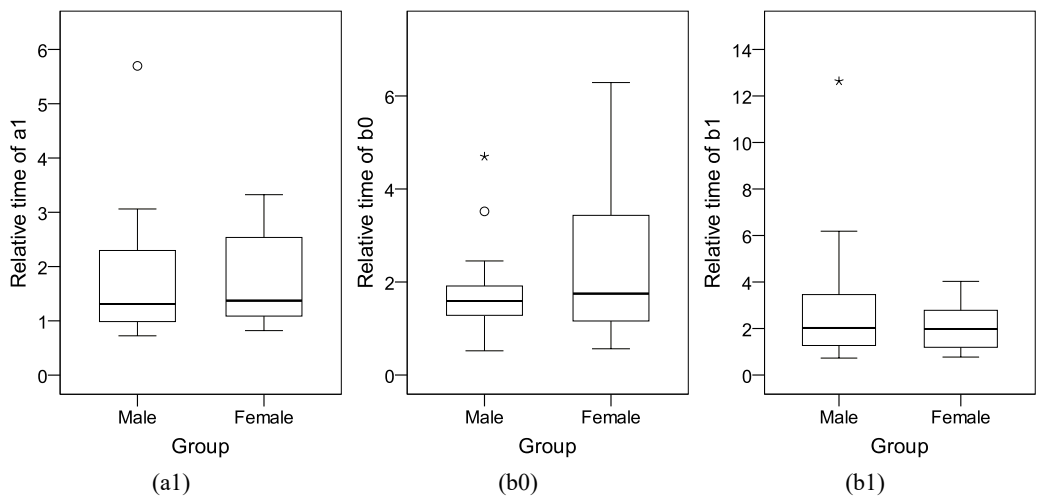

**Figure 4: Relative answering time of each program**

**Table 6. Basic statistics of the number of incorrect answers**

| | | a0 | a1 | b0 | b1 |
|---|---|---|---|---|---|
| Male | Average | 0.29 | 0.53 | 0.41 | 1.24 |
| | Median | 0 | 0 | 0 | 0 |
| | SD | 0.59 | 0.87 | 0.80 | 2.22 |
| Female | Average | 0.38 | 0.63 | 0.94 | 0.75 |
| | Median | 0 | 0 | 0 | 0.5 |
| | SD | 0.50 | 1.26 | 1.48 | 1.06 |
| | p-value | 0.64 | 0.96 | 0.47 | 0.93 |

**Answer to RQ3**: Next, to answer RQ3, we set program a0 as the benchmark, because it had the lowest memory requirement. We focused on the ratio of reading time for program a0 as compared to the other programs. That is, we divided the answering time of each program by the answering time for program a0. A large value indicated a long answering time as compared to the benchmark, program a0. The results are detailed in Table 5. When focusing on the average, the b0/a0, i.e., relative reading time of b0, that of the male group was smaller, while others were larger than the female group. When focusing on the median, the b1/a0 of female group was smaller, and others were larger than the male group. The differences were not large, nor were they statistically significant on 0.05 level. See the bottom row of Table 5.

Distributions of the relative time of each group are shown in Figure 4 using boxplots. Overall, the distributions of answering time did not differ substantially among the programs, except for program b0. Thus, the answer to RQ3 is "No. There is no explicit difference in the relative understanding speed between programs needing much memory and those needing less memory."

**Correctness of the answers**: We also considered the correctness of the answers. Basic statistics of the number of incorrect answers are shown in Table 6. The averages between the two groups were similar. The median of the number was 0, except for program b1 among female subjects; and the differences were not statistically significant at 0.05 level (see the bottom row of table). Therefore, the correctness of the answers was deemed similar between the groups.

## 5 Discussion

### 5.1 How to interpret and utilize the results?

Although past studies such as [32] showed that memory performance is different between males and females in multiple aspects, this experiment demonstrated that the code reading speeds of females were similar to those of males. This might be because code reading is complex task that requires logical thought and other attributes in addition to memory performance. In the female group, the reading time of program b0 was longer than that of b1. When subjects were asked for a reason for this performance, some of them replied that they were confused by calculation statements (i.e., a = a * 2 and b = b + 6) between 'if' statements in program b0. Therefore, we posit that code reading time for female subjects may be affected by structure of programs.

All subjects belonged to the same university, and therefore educational levels were similar. The results may suggest that educational level is more important than developer gender on code reading. Additionally, the performance of developers differs drastically among individuals. For example, approximately half the female subjects read the code faster than approximately half of male subjects for program a1. This is because the median of code reading time of a1 was almost same between male and female groups.

Past studies [4][7] showed that female students and developers has less confidence than males, and therefore they may have stereotype threat explained in section 1. If so, our result is effective to suppress such stereotype. Based on past studies [3][31], if we describe that educational level is more important than developer gender on code reading, code reading speed of female developers may be improved. Additionally, suppressing such stereotype is expected to increase the choice of careers available to female professionals [6].

### 5.2 Threats to validity

Subjects in our experiment were not professional software developers. Past studies have shown that students can be used instead of professionals in experiments because the differences between students and professionals tend to be relatively small [17][29]. Therefore, we expect that these results will not be meaningfully different when professionals join our experiment. Recruiting professional developers is an area for future work.

The number of subjects was not exceptionally large. Subjective experiments are time consuming, and therefore it is not easy to recruit subjects. For example, there were only ten subjects in the study [24], and the number of subjects was 14 in study [30]. However, we acknowledge that we should increase the number of subjects to enhance the reliability of our results. Additionally, because subjects self-selected on a volunteer basis, there may have been selection bias because they were not randomly selected from the larger student population. However, it is difficult to avoid such bias in this type of experiment. For example, study [29] also acknowledges the existence of such bias. Similarly, Fucci et al. [13] analyzed whether sleep deprivation impacts the developers' performance or not without random assignment (i.e., quasi-experiment).

In our experiment, we used small programs. This is because we focus on the influence of memory performance, as did past studies such as [25], which also relied on short programs. If we used more realistic programs, this would introduce many new factors that would also affect the results, and that would obscure the specific influence of the memory performance on reading time. Understanding a short method in a program is similar to our experiment. There are studies which use small programs in experiments. For example, study [12] used code snippets of four lines to analyze code comprehension. Similarly, study [30] uses code snippets. However, we must consider size when interpreting the results.

## 6 Conclusion

In this study, we focused on developer gender based on the assumption that it affects the performance of developers. In the analysis, we presumed that the ability of developers affected by gender is memory. We analyzed whether understanding speed is different between 17 male and 16 female subjects (i.e., 33 subjects), especially when they read programs requiring much memory. Experimental results showed the followings:

- Overall, code reading speed showed little difference between the female and male groups.
- Even when subjects read programs requiring much memory to understand, code reading speed did not differ substantially between the female and male groups.
- When code reading speed was normalized for each subject, there was no meaningful difference in code reading speed between male and female developers.

Note that the goal of our study is to suppress the influence of the gender biases. Our result suggests that developers' gender is not important factor to code reading speed. The result will be effective to increase the choice of careers available to female professionals, and to suppress stereotype threat explained in section 1. As future work, we will identify other abilities of developers which may be affected by gender.

## REFERENCES

[1] T. Barnes, S. Berenson, and M. Vouk (2006). Participation of Women in Information Technology. In E. Trauth (Ed.), Encyclopedia of Gender and Information Technology, 976-982, IGI Global.

[2] L. Beckwith, M. Burnett, S. Wiedenbeck, C. Cook, and S. Sorte and M. Hastings (2005). Effectiveness of end-user debugging software features: are there gender issues? Proc. of SIGCHI Conference on Human Factors in Computing Systems (CHI), 869-878.

[3] A. Bell, S. Spencer, E. Iserman, and C. Logel (2003). Stereotype Threat and Women's Performance in Engineering. Journal of Engineering Education, 92(4): 307-312.

[4] S. Beyer, K. Rynes, J. Perrault, K. Hay, and S. Haller (2003). Gender differences in computer science students, Proc. of 34th SIGCSE technical symposium on Computer science education (SIGCSE), 49-53.

[5] J. Blackburn, G. Scudder and L. Wassenhove (1996). Improving speed and productivity of software development: a global survey of software developers. IEEE Transactions on Software Engineering, 22(12), 875-885.

[6] V. Borsotti (2018). Barriers to gender diversity in software development education: actionable insights from a danish case study. Proc. of International Conference on Software Engineering: Software Engineering Education and Training (ICSE-SEET), 146-152.

[7] M. Burnett, S. Fleming, S. Iqbal, G. Venolia, V. Rajaram, U. Farooq, V. Grigoreanu, and M. Czerwinski (2010). Gender differences and programming environments: across programming populations, Proc. of International Symposium on Empirical Software Engineering and Measurement (ESEM), (28), 10.

[8] Catalyst (2003). Women in U.S. Corporate Leadership: 2003.

[9] C. Chabris, and M. Glickman (2006). Sex differences in intellectual performance: analysis of a large cohort of competitive chess players. Psychological Science, 17(12), 1040-1046.

[10] S. Dehaghani, and N. Hajrahimi (2013). Which factors affect software projects maintenance cost more? Acta Information Media, 21(1), 63–66.

[11] A. Dunsmore, M. Roper, and M. Wood (2000). The role of comprehension in software inspection. Journal of Systems and Software, 52(2–3), 121-129.

[12] B. Floyd, T. Santander and W. Weimer (2017). Decoding the representation of code in the brain: an fMRI study of code review and expertise, Proc. of International Conference on Software Engineering (ICSE), 175-186.

[13] D. Fucci, G. Scanniello, S. Romano and N. Juristo (2019). Need for Sleep: the Impact of a Night of Sleep Deprivation on Novice Developers' Performance. IEEE Transactions on Software Engineering (published online).

[14] A. Fuentes, and M. Desrocher (2013). The effects of gender on the retrieval of episodic and semantic components of autobiographical memory. Memory, 21(6), 619-632.

[15] E. Gorbacheva, J. Beekhuyzen, J. Brocke, and J. Becker (2018). Directions for research on gender imbalance in the IT profession. European Journal of Information Systems, 28(1), 43-67.

[16] A. Herlitz, L. Nilsson, and L. Bäckman (1997). Gender differences in episodic memory. Memory & Cognition, 25(6), 801-811.

[17] M. Höst, B. Regnell, and C. Wohlin (2000). Using Students as Subjects—A Comparative Study of Students and Professionals in Lead-Time Impact Assessment. Empirical Software Engineering, 5(3), 201-214.

[18] S. Hoogendoorn, H. Oosterbeek, and M. Praag (2013). The Impact of Gender Diversity on the Performance of Business Teams: Evidence from a Field Experiment. Management Science, 59(7), 1514-1528.

[19] T. Ishiguro, H. Igaki, M. Nakamura, A. Monden, and K. Matsumoto (2004). Evaluating the Cost of Program Mental Simulation Based on the Number and Variance of Variable Updates. Technical report of IEICE, SS2004-32, 37-42.

[20] T. Kalil and F. Jahanian (2013). Computer Science is for Everyone! The White House blog. https://obamawhitehouse.archives.gov/blog/2013/12/11/computer-science-everyone

[21] P. Loprinzi, and E. Frith (2018). The Role of Sex in Memory Function: Considerations and Recommendations in the Context of Exercise. Journal of Clinical Medicine, 7(6), article 132.

[22] S. Maitland, A. Herlitz, N. Bäckman, and G. Nilsson (2004). Selective sex differences in declarative memory. Memory & Cognition, 32(7), 1160-1169.

[23] A. May, J. Wachs, and A. Hannák (2019). A. Gender differences in participation and reward on Stack Overflow. Empirical Software Engineering, 24(4), 1997–2019.

[24] S. Müller and T. Fritz (2016). Using (bio)metrics to predict code quality online. Proc. of International Conference on Software Engineering (ICSE), 452-463.

[25] M. Nakamura, A. Monden, H. Satoh, T. Itoh, K. Matsumoto, and Y. Kanzaki (2003). Queue-based Cost Evaluation of Mental Simulation Process in Program Comprehension. Proc. of International Software Metrics Symposium, 351-360.

[26] National Center for Women & Information Technology (NCWIT) (2019). Women and Information Technology by Numbers. https://www.ncwit.org/sites/default/files/resources/btn_05092019_web.pdf

[27] F. Pauls, F. Petermann, and A. Lepach (2013). Gender differences in episodic memory and visual working memory including the effects of age. Memory, 21(7), 857-874.

[28] A. Rayome (2017). CIO Jury: 83% of CIOs struggle to find tech talent. TechRepublic. https://www.techrepublic.com/article/cio-jury-83-of-ciosstruggle-to-find-tech-talent/

[29] I. Salman, A. Misirli, and N. Juristo (2015). Are students representatives of professionals in software engineering experiments? Proc. of International Conference on Software Engineering (ICSE), 666-676.

[30] J. Siegmund, C. Kästner, S. Apel, C. Parnin, A. Bethmann, T. Leich, G. Saake, and A. Brechmann (2014). Understanding understanding source code with functional magnetic resonance imaging. In Proc. of International Conference on Software Engineering (ICSE), 378-389.

[31] S. Spencer, C. Steele, D. Quinn (1999). Stereotype Threat and Women's Math Performance. Journal of Experimental Social Psychology, 35(1), 4-28.

[32] P. Sunderaraman, H. Blumen, D. DeMatteo, Z. Apa, and S. Cosentino (2013). Task demand influences relationships among sex, clustering strategy, and recall: 16-word versus 9-word list learning tests. Cognitive and Behavioral Neurology, 26(2), 78-84.

[33] J. Terrell, A. Kofink, J. Middleton, C. Rainear, E. Murphy-Hill, C. Parnin, and J. Stallings (2017). Gender differences and bias in open source: pull request acceptance of women versus men. PeerJ Computer Science, 3:e111.

[34] Y. Tsugawa, A. Jena, J. Figueroa, E. Orav, D. Blumenthal, and A. Jha (2017). Comparison of Hospital Mortality and Readmission Rates for Medicare Patients Treated by Male vs Female Physicians. JAMA Internal Medicine, 177(2), 206-213.

[35] B. York (2015). The Best Developer Performance Metrics. Medium. https://medium.com/@yupyork/the-best-developer-performance-metrics-6295ea8d87c0